"Some comments on the possible causes of climate change"

L. Padget and J. Dunning-Davies,

Department of Physics,

University of Hull,

Hull HU6 7RX,

England.

email:  j.dunning-davies@hull.ac.uk

**Abstract**

Climate change is an important current issue and there is much debate about the causes and effects. This article examines the changes in our climate, comparing the recent changes with those in the past. There have been changes in temperature, resulting in an average global rise over the last 300 years, as well as widespread melting of snow and ice, and rising global average sea level. There are many theories for the causes of the recent change in the climate, including some natural and some human influenced. The most widely believed cause of the climate change is increasing levels of Greenhouse gases in the atmosphere and as the atmosphere plays an important role in making our planet inhabitable, it is important to understand it in order to protect it. However, there are other theories for the cause of climate change, the Sun and cosmic rays, for example, are felt by some to have a significant role to play. There is also well-established evidence that the three Milankovitch cycles change the amount and alter the distribution of sunlight over the Earth, heating and cooling it. There are many influences on our planet and they all have differing levels of impact. The purpose of this article is to review the present overall position and urge open, reasoned discussion of the problem.

# Introduction

Climate change is quite possibly the most hotly debated issue at the moment and there are many conflicting views about the causes. As well as being an issue that affects most of us on a daily basis, it is a very important political issue.

So what is happening to convince people that there is a problem? The BBC news reports that average global temperatures have risen by 0.7° C over the last 300 years. 0.5 ° C of that warming occurred during the 20[th] century, and most of that occurred between 1910-1940 and from 1976 onwards. Four out of five of the warmest years ever to be recorded were in the 1990's, with 1998 being the warmest year globally since records began in 1861. It is also widely believed that arctic sea ice is thinning and that there has been an average increase of between 0.1 and 0.2 meters in sea levels globally over the last 100 years. In many high and mid level areas in the northern hemisphere, precipitation has increased by 0.5-1% per decade. Finally in Asia and Africa the frequency and intensity of droughts has increased in the last few decades. (BBC08)

The Intergovernmental Panel on Climate Change (IPCC) begins its *Climate Change 2007: Synthesis Report* with the statement "Warming of the climate system is unequivocal, as is now evident from the observations of increases in global average air and ocean temperatures, widespread melting of snow and ice, and rising global average sea level". According to the IPCC: temperature increase is widespread over the globe and greater at higher northern latitudes, with land regions warming faster than the oceans. Global average sea level has risen since 1961 at an average rate of 1.8 [1.3 to 2.3][1] mm/year and since 1993 at an average rate of 3.1 [2.4 to 3.8] mm/year, due to thermal expansion, melting glaciers and icecaps, and the polar ice sheets. It follows that observed decreases in snow and ice extent are also consistent with warming. Satellite data has shown that arctic sea ice has shrunk by 2.7 [2.1 to 3.3] % per decade with larger decreases in summer of 7.4 [5.0 to 9.8] % per decade. Also mountain glaciers and snow cover on average have declined in both hemispheres. They state that there is a *very high confidence* that the earlier timings of spring events and poleward and upward shifts in plant and animal ranges are linked to recent warming. In some marine and freshwater systems, shifts in ranges and changes in algal, plankton and fish abundance are with *high confidence* linked to rising water temperatures as well as ice cover, salinity, oxygen levels and circulation. Of the 29,000 plus observational data series, taken from 75 studies, which show significant change in physical and biological systems, more than 89% are consistent with the change expected as a response to warming. Nevertheless, there is a geological imbalance with a notable lack of data and literature on changes coming from developing countries. (IPCC, 2007)

However, according to reports from the U.S. National Oceanic and Atmospheric Administration (NOAA), almost all the allegedly 'lost' ice is back, they show that ice which had shrunk from 13 million square kilometres in January 2007 to just 4 million in October is almost back to its original level and figures show that there is nearly a third more ice in Antarctica than is usual for this time of year. Scientists are saying that last winter, the northern hemisphere endured its coldest winter in decades and that snow cover across that area was at its greatest since 1966. One exception to this was Western Europe which experienced unseasonable warm weather; the UK reported one of its warmest winters on record. Vast parts of the world have, however, suffered chaos because of some of the heaviest snowfalls in decades, including central and southern China, the United States and Canada. In China, the snowfall was so severe that over 100,000 houses collapsed under the weight of it. Jerusalem, Damascus, Amman and Saudi Arabia all reported snow and below zero temperature and in Afghanistan snow and freezing weather killed 120 people. (Bonnici, 18.02.08) (Brennan, 19.02.08)

It is clear that our planet's climate is changing and some opinion suggests that this is down to us. But some scientists believe that because the climate has changed naturally before that it is supposed to change and therefore disagree that there is even a problem. Our climate is an incredibly complex system and there is doubt over whether enough is known about it to make predictions and whether the computer models that are being used are adequate. (BBC08)

**Possible causes.**

The atmosphere that surrounds the Earth plays an essential role in making our planet habitable, it is transparent to the visible radiation emitted by the Sun and this heats the Earth's surface, without it the temperature would soar by day and plummet by night, and the average temperature would be around -18°C. About 30% of the Sunlight that reaches our planet is reflected back to space by clouds, dust or the ground, more than 20% is absorbed in the atmosphere and almost 50% is absorbed by the Earth's surface. Some of the infrared radiation that is radiated by the Earth's Sun-warmed surface escapes through the atmosphere directly into space but most of it is absorbed on the way by clouds and greenhouse gases which release part of that heat into space and radiate some back to the surface increasing the temperature in the lower atmosphere. As the temperature of the Earth's surface rises, the amount of IR radiation increases. The temperature adjusts until a delicate balance is achieved. Unlike the two main components of air, oxygen (20%) and nitrogen (78%) that have a linear diatomic structure, greenhouse gases in the atmosphere such as water, carbon dioxide and methane have three or more atoms which make them well suited to absorbing radiation. As these greenhouse gases accumulate they block each other's radiation to space and so, the more greenhouse gas there is, the warmer it gets. The average height at which the radiation can now escape to space then begins to increase and at higher altitudes the temperatures are cooler and radiation into space decreases. The system then starts to readjust, more water evaporates from oceans and lakes and sea ice which reflects Sunlight back into space begins to melt, reducing the reflection, these both amplify the warming effect. (Henson, 2006)

Another theory for the cause of global warming, developed by Henrik Svensmark amongst others, is that the Sun and cosmic rays play a role in the change in our climate. They believe that cosmic rays are an essential ingredient, which experts have so far been slow to appreciate. The cosmic rays must break through three defensive shields before they can reach the Earth's surface, first the Sun's magnetic field then the Earth's magnetic field and finally the air around us and only the most energetically charged particles can get as far as sea level. In Svensmark's theory it is these energetically charged particles called muons or heavy electrons, which are produced when cosmic rays hit the atmosphere, that help clouds to form low in the air and cool the Earth. Whist some clouds higher up can have a warming effect; these clouds which are less than 3000 meters high keep the Earth cool. Put simply this means more cosmic rays, more clouds and cooler temperatures. (Svensmark, et al., 2007)

The clouds play a very important role in our climate, about 60% of the globe is covered by cloud and we all appreciate how important cloudiness is in determining the temperature on a day to day basis. Clouds modulate the Earth's radiation balance, both in the visible and infrared spectra. Clouds cool the Earth by reflecting incoming Sunlight, the tiny drops in clouds can scatter between 20 and 90% of the Sunlight that reaches them, a cloud free Earth would absorb nearly 20% more heat from the Sun that it does at present. However, the clouds also have a warming effect on the Earth, they absorb the infrared radiation emitted from the surface and reradiate it back down. This process traps heat like a blanket and slows the rate at which the surface cools down. The clouds reflect about 50$Wm^{-2}$ of solar radiation up to space and radiate around 30$Wm^{-2}$ down to the ground, the net effect being 20$Wm^{-2}$ cooling on average. This greatly exceeds the 4$Wm^{-2}$ warming due to atmospheric carbon dioxide levels doubling from 300 to 600 ppm. What we don't know however, is what the net cooling or warming effect of all clouds on Earth will be in a changing atmosphere or how the clouds themselves will be changed by a change in the temperature of the Earth. If the cooling effect of clouds increases more than the heating effect does, the clouds would reduce the magnitude of the greenhouse-induced warming but speed its arrival, this is called negative feedback. Both effects decreasing could have the same effect but if the cooling decreases more than the heating the cloud changes would boost the magnitude of the warming but delay its arrival. In any scenario the important factor is the net effect of the clouds. To complicate matters however, the altitude of the clouds has an influence. High clouds have a net warming effect, they block little incoming solar radiation but because they are at low temperature they return little outgoing infrared radiation to the Earth's surface. Clouds at a low altitude have a net cooling effect because they have a high albedo and being at a temperature which is nearly as warm as the surface of the Earth they emit nearly as much infrared radiation to space as the surface would under clear skies. (Rossow, et al., January 1995)

There is also well established evidence that the Three Milankovitch Cycles in the Earth's rotation and orbit, change the amount and alter the distribution of Sunlight over the Earth, heating and cooling the Earth over cycles of 100,000- 41,000 and 23,000 years. (Page, 27.06.07) The Milankovitch cycles is the name given to the collective effects of changes in the Earth's movements on the climate. The eccentricity, axial tilt and precession of the Earth's orbit vary in several patterns which have resulted in 100,000 year ice age cycles over the last few million years. The Earth's tilt goes up and down ranging from about 21.8° to 24.4° and back over a approximately 41,000 year cycle. The tilt is currently around 23.4° and decreasing. When the tilt is most pronounced it gives rise to stronger summer Sun and weaker winter Sun. Ice ages often occur because as the tilt decreases the progressively cooler summers cannot melt the past winter's snow. The Earth's orbit around the Sun is not precisely circular but elliptical in shape, with the Sun positioned slightly to one side of the centre point. The eccentricity, or 'off-centeredness' of the orbit varies over time in a complicated way, the result is two main cycles, one averages about 100,000 years long and the other 400,000 years. When the eccentricity is low there is little change through the year in the distance between the Earth and Sun. When the eccentricity is high the Sunlight reaching the Earth can be more than 20% stronger at perihelion (currently January) than at aphelion (currently July). The Earth's axis also plays a role, the main cycle of the rotation around the axis is known as the precession, and takes about 26,000 years. It shifts the dates of the perihelion and the aphelion forward by about one day every 70 years. Currently the Earth is 3% closer to the Sun in early January (perihelion) than in early July (aphelion) with about 7% more solar energy reaching the Earth at perihelion. In about 13,000 years the Earth will be closest to the Sun in July instead of January; this will intensify the seasonal changes in solar energy across the Northern Hemisphere and weaken them in the South.

## $CO_2$ emissions

It is reported that man is producing and releasing into the atmosphere higher levels of carbon dioxide through increased industry and de-forestation and that this is affecting our climate by adding to the layer of greenhouse gases in the atmosphere. However, there are many factors which affect the levels of greenhouse gases in the atmosphere; the sea for example absorbs carbon dioxide.

Geologist Dr Norman J Page published an article in June 2007 entitled 'Climate Change and Carbon Dioxide' for the Alpha Institute for Advanced Study saying "As a geologist, I find the current climate of fear in which the debate on Global Warming is conducted very alarming". He starts by pointing out some common misconceptions often reported in the media:

The United States is often referred to as the worlds biggest polluter but, whilst the U.S. does emit a large amount of $CO_2$, the land use patterns means that they also absorbs a large amount and it is the net amount not the amount emitted which is the important figure. It has been shown by a paper published in Science Magazine in 1998 that, when the net amount of carbon dioxide is taken into account, North America actually takes up more carbon dioxide than it emits by about 100 million tonnes per year, whilst Europe emits a large amount overall. Carbon dioxide is often reported as being the biggest offender but in fact water vapour is Earth's most abundant greenhouse gas and $CO_2$ makes up less than 3%. Dr Page states that the Earth is now impoverished in $CO_2$ and that at various times in the last 550 million years levels of CO2 have often been four or five times the current levels and at times ten to fifteen times greater.

Annually, human contribution to greenhouse gas is said to be between one and two tenths of a percent and it has been suggested that termites alone produce ten times more greenhouse gas than humans. It has also been reported that the amount of $CO_2$ produced by the population of India breathing is more than all of the coal burning plants in the U.S. Page states "If we eliminated human use of fossil fuels entirely it would have little impact on future temperatures". Dr Roth, at MIT, has shown that over time scales of more than 10 million years it is very difficult to prove a connection between the climate and $CO_2$. There is, however, a connection between temperature and $CO_2$ over shorter time intervals but data extracted from ice cores show that a natural warming period precedes a $CO_2$ increase and is not caused by it. (Page, 27.06.07)

The latest data from Britain's Climate Research Unit shows that the global mean temperature in October 2007 was 0.159 degrees lower than in October 1997; there has been no warming in the last 10 years but in this same period the levels of $CO_2$ have increased by 6%. The latest data from the Met Office also shows that the average global temperature for the first quarter of 2008 was cooler than the average of any year since 1996, whilst levels of $CO_2$ have risen by 6%. (Page, 14.04.08). On the other hand the IPCC state that eleven of the last twelve years, from 1995 to 2006, rank among the twelve warmest years in the instrumental record of global surface temperature since 1850.

In November 2007, the Parliamentary Office of Science and Technology periodical Postnote published the report *Climate Change Science*. It stated that the atmospheric concentration of $CO_2$ has increased from a pre-industrial concentration of about 280 parts per million (ppm) to 379 ppm in 2005 and that over the last 650,000 years $CO_2$ varied within a range of 180 to 300 ppm, but there were approximately 90,000 measurements of $CO_2$ levels made since 1812, through the $19^{th}$ and early $20^{th}$ century which show that levels of $CO_2$ were at about 440 ppm in the 1820s and 1940s and about 370 ppm in the 1850s. The historical chemical data shows a clear trend, with the changes in $CO_2$ tracking the changes in temperature and therefore the climate. These measurements were made by chemists, several of whom had Nobel Prize level distinction, and the chemical methods used to make the measurements are good enough to measure with high accuracy, for example the Pettenkofer process was the standard analytical method for determining atmospheric carbon dioxide levels between 1857 and 1958 and usually achieved an accuracy better than 3%. This data was recently published by Ernest-Georg Beck, but modern climatologists have generally ignored the historic figures, discrediting the methods as unreliable despite the techniques being standard textbook procedures in several different disciplines. Furthermore this data was recently not acknowledged by the IPCC when it published its findings. (Beck, March 2007)

The factors that drive climate change are separated into two categories, forcings and feedbacks. Changes in solar energy output or in the concentration of greenhouse gases, natural or otherwise, are classed as forcings. Scientists quantify and compare the contributions of different agents that affect surface temperatures by measuring their 'radiative forcing' which can be positive or negative and is measured in Watts per metre squared. Feedbacks are internal climate processes that amplify or reduce the climate's response depending on how responsive the climate is to various forcing processes, for example, a warmer atmosphere can hold more moisture which itself acts as a greenhouse gas causing further warming, this is a positive feedback.

Postnote reports that scientists have estimated the combined human-caused radiative forcing to be +1.6Wm$^{-2}$, and have decided that it is *extremely likely*[2] that humans have exerted a substantial warming influence on the climate, and that this radiative forcing is *likely* to be at least five times greater than the radiative forcing due to solar changes. Solar irradiance is estimated to have caused a small warming effect +0.12Wm$^{-2}$. It concludes that, for the period from 1950 to 2005, it is extremely unlikely that the combined natural radiative forcing (solar plus volcanic sources of aerosols) has had a warming influence that is comparable with that of the combined human made radiative forcing. But there is an uncertainty in the predictions of the future climate states that is due to the uncertainty in the magnitude of climate forcings and feedbacks. The IPCC has classed the level of scientific understanding (LOSU) as low for the climate forcings of Solar irradiance, contrails from aircrafts, cloud reflectivity and water vapour from the methane in the upper atmosphere.

Scientists use climate models to help quantify these forcings and feedbacks to predict the future state of the climate. Mathematical equations of the climate are fed into a three dimensional grid of points that cover the Earth's atmosphere and oceans. The resolution of the model, which is essentially the spacing between the points on the grid, is limited by the power of the computer used by the researcher, but is generally less than 150km in the horizontal direction and 1km in the vertical, with finer resolution near the surface of the Earth. Some scientists are confident that climate models provide credible estimates of future climate change at least on large scales, because their design is based on established physical laws which are used in weather forecasting models. The models have also been used to reproduce features of past climates and climate changes. But these models do have their limitations, some processes such as cloud formation, occur on time and space scales which are too small for the climate models to resolve. Modellers deal with this problem by representing small scale processes with average values over one grid box, but these assumptions are often a big source of climate model uncertainty. Climate modellers run a number of programs and average the results, hoping to try and remove or reduce some of the effects of natural variability, leaving the human caused changes. (Smith, November 2007) (Page, 29.11.07) (Page, 14.04.08)

The Intergovernmental Panel on Climate Change is a United Nations organisation and is there to help policymakers decide how to respond to climate change. Its role is not to carry out any scientific investigation but to evaluate studies carried out by thousands of researchers around the world and then synthesise the results into one report. For the 2001 and 2007 reports there were three working groups to deal with: the basis in physical science, impacts, adaptation and vulnerability and mitigation, each group is headed by a pair of scientists, one from a developed and one from a developing country. Climate Change 2007: Synthesis Report is based on the assessment carried out by the three working groups. The report starts by discussing the observed changes in the climate and their effects, they say it is *very likely* that over the past 50 years hot days and hot nights have become more frequent over most land areas and cold days, cold nights and frost have become less frequent. From observational evidence from all continents and most oceans it seems that many natural systems are being affected by regional climate changes, particularly temperature increases.

The report then goes on to discuss the causes of the change. The energy balance of the climate system is altered by changes in the atmospheric concentrations of greenhouse gases and aerosols, land cover and solar radiation. Levels of global greenhouse emissions due to human activity have grown since pre-industrial times and there was an increase of 70% between 1970 and 2004. The annual emissions of carbon dioxide grew by approximately 80% between 1970 and 2004.

The atmospheric concentration of $CO_2$, 379 ppm, far exceeds the natural range over the last 650,000 years and this increase is primarily attributed to fossil fuel use. At 1774 ppb the atmospheric concentration of methane also exceeds the natural range over the same time period and it is *very likely* that this is predominantly due to agriculture and fossil fuel use. The IPCC concludes that there is a *very high confidence* that the net effect of human activities since 1750 has been one of warming and that continued greenhouse gas emissions at the current rate or above would cause further warming and induce changes in the global climate system that would *very likely* be larger than those observed already.

It is often reported that there is a consensus of opinion amongst the leading climate scientists that greenhouse gases are the cause of climate change but this is not the case. According to many involved with the IPCC including Professor John Christy, lead author, not all of the 2500 plus top scientist listed as contributors agree with findings of the report, and some have had to fight to have their names removed. There have even been claims that the IPCC has censored its scientists. Professor Frederick Seitz wrote a letter to the Wall Street Journal saying that the version of the IPCC's latest report that was released was not the version approved by the scientists listed on the title page. He claims that at least 15 of the key sections in the science chapter have been deleted including the following statements "None of the studies cited above has shown any clear evidence that we can attribute the observed [climate] changes to the specific cause of increases in greenhouse gases" and "No study to date has positively attributed all or part [of the climate change observed to date] to anthropogenic causes". The IPCC did not deny removing any sections but said there was "no bias" in their report. (Durkin, 2007)

American politician Al Gore is a passionate believer that greenhouse gases are the cause of climate change and has been campaigning world wide for change. The documentary *An Inconvenient truth* comprises the lectures he has been giving around the world on climate change and interviews with Gore. He claims that there is no doubt that the $CO_2$ emitted by humans thickens the layer of greenhouses gases and that this warms the planet. However, the documentary focuses more on the effects that global warming is having on the climate and the weather than showing that it is caused by $CO_2$. Gore says that the idea the worlds temperature has been warmer than it is at present is untrue and shows a graph which depicts the temperature during the Medieval Warming Period as being much cooler than that of today. He discusses the ice core method used to measure levels of temperature and carbon dioxide at periods in the past but does not mention what conclusions have been drawn from the results obtained. Gore then uses a graph showing the levels of $CO_2$, going back millions of years which shows that levels have never exceeded 300 ppm until now, which does not agree with the data published by Beck. According to *An Inconvenient Truth,* the poles experience a much greater impact due to global warming, claiming that because Earth's climate is non-linear, a global rise of 5°F would cause a 1°F rise at the equator but a 12°F rise at the poles. During the *Inconvenient truth lecture* Gore uses the images of icebergs breaking off and falling into the sea, ice sheets melting and a polar bear clinging to a melting iceberg; all images which have widely been used to show the need to fight climate change. It is claimed that our global warming is melting icebergs which is causing polar bears to drown in the oceans. However, he fails to mention that the polar bear population has soared in recent years or even that many of the photographs were taken in August when melting is normal. Gore's solution to the problem of global warming is to reduce carbon emissions by such measures as using renewable energy and increasing both electricity and transport efficiency. Whilst the message of the documentary is conveyed strongly and emotively, using footage of natural disasters and stories of his own personal tragedies, the scientific content of the documentary seems questionable, many of the graphs shown do not have axes or scales and much of the data quoted has no units. (Gore, 2007) (Brennan, 19.02.08)

In response to *an inconvenient truth,* Martin Durkin made the documentary The Great Global Warming Swindle, which was shown last year on Channel Four. It features interviews and opinions of many top climate scientists who disagree with the greenhouse gas explanation of climate change. The documentary begins by discussing a few of the points in history where the temperature was significantly higher than that of today, the Medieval Warm Period, before the Little Ice Age for example and the Holocene Climatic Maximum, when the temperature was significantly higher than it is now for three millennia. The warming event seems to have peaked between 11,000 and 8,000 years ago and North-western Europe experienced warming, while there was cooling in the south. There is evidence at 120 of 140 sites across the western Arctic of temperatures warmer than at present. At 16 sites where quantitative estimates were obtained they show local temperatures that were on average 1.6±0.8 °C higher than present. The cause of this event is believed to be the Milankovitch cycles and a continuation of changes that caused the end of the last glacial period. When the axial tilt was at 24° and the Earth at its nearest approach to the Sun the warming would have been at its maximum in the Northern Hemisphere. The Milankovitch forcings would have provided 8% more solar radiation, calculated to be +40W/m² to the Northern Hemisphere in the summer, causing greater heating. (Durkin, 2007) (Davis, et al., 2003) (Kaufman, et al., 2004)

The data the IPCC used to demonstrate climate change, as well as that used by Al Gore and many others, indicates a period of intense warming during the early 20th century. Clearly in the period between 1905 and 1940 industry was fairly primitive. However, in the years after the 2nd World War when the world economy boomed, industry thrived and $CO_2$ levels soared but according to the data, from approximately 1940 to 1960 the world cooled. So it seems that the time does not fit with greenhouse warming.

It is known that incoming radiation from the Sun is trapped by the greenhouses gases in the troposphere, so it follows that, if an increasing level of greenhouses gases is responsible for the warming, the rate of warming should increase the higher you go. The temperature of the atmosphere can be measured using a satellite or by weather balloon. Using both methods it has been found the rate of warming is in fact higher at the surface than the upper atmosphere, which does not fit the theory.

It is often reported in the media that a temperature rise of just a few degrees could have a huge warming impact, melting the icecaps. But records show that Greenland has been much warmer than it is today and did not have a big warming event. Pictures of pieces of ice breaking off and falling into the sea are also often shown, but the icecaps are always naturally expanding and contracting and it is perfectly normal for pieces to break off.

Dr Ian Clarke used ice drilling to try and find out if there is a link between $CO_2$ and the climate and although he found a connection it was an unexpected one. The temperature on Earth leads the levels of $CO_2$ by 800 years. Several major ice surveys since have confirmed these findings. This can be explained by looking at the oceans. They contain $CO_2$ and each year they absorb some and emit some, but how much depends on the temperature. When the temperature is warm they release more and when it is cooler they absorb more. The oceans are so big that it takes a long time for them to warm or cool, often hundreds of years, which explains the time lag between the temperature and level of $CO_2$.

There is also doubt raised in the documentary about the accuracy of the climate models used. A model is only as good as the assumptions on which it is based. These climate models are based on hundreds of assumptions and it only takes one to seriously distort a models findings. The other concern with the current climate models is that they assume $CO_2$ is the main climate driver and do not incorporate any other possible influences. It is claimed that tweaking the parameters of the model, even slightly, can show a number of possible outcomes.

$CO_2$ makes up 0.054% of all the gases in the atmosphere, the percentage of anthropogenic $CO_2$ is even smaller and it is known that $CO_2$ accounts for just 3% of greenhouse gases which are themselves only a small part of the Earth's whole climate system. So what do the greenhouse sceptics believe is driving the changing climate? The Sun!

Sun spots are intense magnetic fields which appear at times of increased solar activity but even before this was understood astronomers would count the number of Sun spots in the belief that more heralded warmer weather. Edward Maunder noticed in 1893, during the Little Ice Age that there were barely any Sun spots visible, this became known as the Maunder Minimum. Eigil Friis-Christensen compared the Sun spots with temperature over the last 120 years and found a very close correlation, using astronomical data for the past 400 years the comparison was taken further back and was found to be intimately linked.

Astrophysicists from Harvard University conducted a study, in 2005, into the temperature-carbon dioxide and temperature- Sun relationships. No obvious link was found linking temperature and carbon dioxide, although a close link on a decade to decade basis was found between temperature and the Sun; this was based on independent data from NASA and the U.S Oceanic and Atmospheric Administration, displays. (Durkin, 2007)

## Cosmic Rays

In February 2007 A&G magazine- News and reviews in Astronomy & Geophysics published the article *Cosmoclimatology: a new theory emerges* by Henrik Svensmark. This report contains a lot of in depth information on the research finding by Svensmark and emerging scientific evidence. The Sun's influence on our climate had long been recognised, by Herschel in 1801, Eddy in 1976, Friis Christensen and Lassen in 1991, but Henrik Svensmark and Eigil Friis Christensen noted the link between this relationship and the small 0.1% variations in the solar irradiance over a solar cycle measured by satellites. The pair published a paper titled "Variation of cosmic ray flux and global cloud coverage- a missing link in solar-climate relationships" in 1997 after they announced their findings at the COSPAR space science meeting in Birmingham in 1996. (Svensmark, February 2007)

*The Chilling Stars- A new theory of climate change* was published in 2007 by Henrik Svensmark and Nigel Calder. The theory that cosmic rays play a role in the change in our climate has recently been developed and this book goes into detail to explain the theory and the evidence for it. It stems from Svensmark's research at the Danish National Space Centre. The book is a collaboration between Svensmark, who contributed the majority of the scientific input, and Calder who wrote it up. The pair met in 1996 when they were introduced by Friis-Christensen, who has also contributed toward the cosmic ray theory. They continued to discuss the theory over the following years and decided to collaborate on a book. The text evolved in 2005/06 whilst the intensive research continued, despite difficulties with funding.

Our Ancestors sometimes believed that the Moon and stars sucked heat from the Earth, but astronomers now know that most of the brightest stars are far hotter than the Sun. Despite this Svensmark and Calder believe that, when the biggest of the stars expire in mighty supernova explosions and spray the galaxy with cosmic rays, they do in fact cool the Earth by making the atmosphere cloudier.

When cosmic rays were detected by an Australian scientist nearly a century ago, it seemed that they were an interesting but unimportant extra, but it could be that they are an essential ingredient in the universe and a vital component in changing the climate on our planet.

Svensmark saw the first clues that cosmic rays have an effect on the climate when he looked at the alternating episodes of warmth and cold over the past few thousand years, starting with the Little Ice Age which peaked around 300 years ago, giving way to the present warm interlude. At the time of the Little Ice Age the Sun was in an unusual state, the Maunder Minimum, and there was very low Sun spot activity. This coupled with a jump in production rate of radiocarbon atoms and other long lived tracers, which are made by cosmic rays in nuclear reactions in the air, is an indication of low magnetic activity. Cosmic rays are deflected away from Earth by the Sun's magnetic field, but when it weakens more of them can reach the Earth. Since the most recent ice age ended 11,500 years ago, there have been nine chilling events like the Little Ice Age and these have always been associated with high counts of radiocarbons and other tracers.

The Cosmic rays must break through three defensive shields before they can reach the Earth's surface. First the Sun's magnetic field, then the Earth's magnetic field and finally the air around us. Only the most energetically charged particles can get as far as sea level. In Svensmark's theory it is these energetically charged particles called muons or heavy electrons, which are produced when cosmic rays hit the atmosphere, that help clouds to form low in the air and cool the Earth. Whilst some clouds higher up can have a warming effect; those clouds which are less than 3000 meters high keep the Earth cool. Put simply this means more cosmic rays, more clouds and cooler temperatures. During the 20$^{th}$ century the Sun's magnetic shield more than doubled in strength and so reduced the cosmic rays and clouds enough to explain a large fraction of the global warming reported by climate scientists. When they first revealed their ideas about the link between cosmic rays, clouds and the climate, Svensmark and his colleagues experienced a lot of criticism. To gain credibility for their theory, the team had to find out exactly how the cosmic rays affect the formation of clouds. Understanding of where clouds came from was surprisingly limited. Elementary text books said that when air becomes cold enough, moisture can condense and form clouds. But there must first be small specks floating in the air, the cloud condensation nuclei on which the water droplets can form. They needed to be seeded too, but how that happened was a mystery. The experiment SKY was set up in 2005 at the Danish National Space Centre and began to provide the scientists with some answers. Cosmic rays enter through the laboratory ceiling and into a large box of air, releasing electrons in the air which then encouraged the clumping of molecules to make micro-specks. The micro-specks are capable of gathering into the larger specks which are needed for the formation of clouds, the speed and efficiency at which the electrons worked took the team by surprise. In 2006 a more elaborate experiment CLOUD was set up at CERN, Europe's particle physics lab in Geneva, using accelerated particles to simulate the cosmic rays and test other possible effects. The influx of cosmic rays on the Earth depends not just on the state but on where we are in the galaxy. The Sun, along with the Earth, orbits round the centre of the Milky Way and sometimes finds itself in a dark region where hot, bright explosive stars are few. In those regions cosmic rays are relatively scarce and the Earth's climate is warm. This is referred to by geologists as the hothouse mode. In the opposite periods when the starlight and cosmic rays are abundant, the planet goes into an icehouse phase and ice sheets and glaciers form. An Israeli scientist has suggested that the major changes in theses phases could be accounted for by visits to the bright spiral arms of the Milky Way. (Svensmark, et al., 2007)

Although we do not fully understand the Sun- climate interface it is seems that Sun spot activity is a good proxy for solar activity in general. The 11 year Sun cycle is well known but other cycles of varying lengths have been suggested. One which is of particular interest is an approximately 1000 year cycle which is believed to have peaked recently. A previous peak in this cycle is reported to have produced the well documented medieval warming period. The forecast for the next two solar cycles, number 24 and particularly number 25, show that we are heading for a period of reduced activity for at least the next 15-20 years, perhaps until the mid century and this would produce generally declining temperatures, as opposed to the temperature rises predicted by the IPCC. (Page, 29.11.07)

Recently Richard Black, BBC environment correspondent wrote a report for BBC news online entitled *'No Sun link' to climate change.* This article is about a new scientific study carried out by Dr Mike Lockwood of the UK's Rutherford-Appleton Laboratory and Dr Claus Froehlich from the World Radiation Centre in Switzerland. Their findings claim to disprove the cosmic ray hypothesis developed by, amongst others, Henrik Svensmark. The study was initiated by Dr Lockwood, partly in response to the television documentary *'The Great Global Warming Swindle'*, which features the cosmic ray hypothesis. He claims that "All the graphs they showed stopped in about 1980, and I knew why, because things diverged after that" "You can't just ignore bits of data that you don't like". That is simply not true, the graphs extend to at least 2000. See figures 5 and 6, taken from the 2007 Astrophysics & Geophysics article, for examples of just 4 graphs which continue past 1980.

The scientists looked at the solar output and cosmic ray intensity over the last 30-40 years and compared those trends with the graph for global average surface temperature showing a rise of about 0.4°C over the period. This article does not go into a great deal of scientific detail to explain the findings of this new study, to support the claims there are two graphs, one showing the cosmic ray count and one showing the global mean surface air temperature from 1975 to 2005. As discussed before, the Sun varies on a cycle of approximately 11 years between periods of high and low activity, but that cycle coinciding with longer term trends saw most of the 20$^{th}$ century showing a slight but steady increase in solar output. All except for the period between 1985-90 when the trend appears to reverse and the solar output declines. However, the other graph shows that during this period the temperature rises just as fast. But although the graph does not show any obvious link, it is quite possible that it is part of a long term trend as shown by Lassen and Friis-Christensen in figure 5.

The study has been criticised by some for not recognising the work of Svensmark, Friis-Christensen and others. The article concludes that "changes in the Sun's output cannot be causing modern-day climate change". Lockwood does agree that there is a cosmic ray effect on cloud cover but thinks that, whilst it may have had a significant effect on the climate of pre-industrial Britain, it cannot be applied today because the situation is completely different. Lockwood's analysis is said by Black to have "put a large, probably fatal nail" into the cosmic ray theory. (Black, 10.07.07)

In a follow up article, also entitled *'No Sun link' to climate change,* Black discusses the work done by scientists at Lancaster University to further contradict the cosmic ray theory. The team, headed by Professor Terry Sloan, has found that there has been no significant link between solar activity and cosmic ray intensity in the last 20 years. They presented their findings in the Institute of Physics journal and explained that they used three different methods to search for a correlation but found virtually none. To try and establish a link, Professor Sloan's team looked for periods in time and for places on Earth where weak or strong cosmic ray arrivals had been documented and then examined whether that had affected the cloudiness observed in those locations or at those times.

When speaking to the BBC News, Sloan explained that the Sun sometimes throws out a huge burst of charged particles, described as a 'burp', and that they had looked to see if cloud cover had increased after one of these bursts and that they found nothing. They did observe what they described as a 'weak' correlation between cosmic ray intensity and cloud cover over the course of one of the Sun's natural 11 year cycle, but concluded that at most, cosmic ray variability could account for only a quarter of the changes in cloudiness.

Dr Giles Harrison of Reading University, who is a leading researcher in the physics of clouds, has also conducted research, looking at the UK only, which has suggested that cosmic rays only make a very weak contribution to cloud formation. Sloan says that the team did try to corroborate Svensmark's hypothesis but was unable to and the article concludes that the assessments made and the conclusions reached by the Intergovernmental Panel on Climate Change last year are correct and that Svensmark has no reason to challenge them. (Black, 03.04.08)

## Conclusion

It is clear that there are many different theories for the cause of climate change and valid evidence to support them all. It seems that a great many politicians have made up their minds that greenhouse gases are the main driver of climate change, and they do not seem very willing to explore other possibilities. The media seems to have followed their lead. They report greenhouses gases being the cause of climate change as irrefutable fact and rarely report on any other possibilities. With sources of fossil fuels becoming more unstable because of dwindling stocks and political issues, warming due to carbon dioxide emissions gives even more reason to switch to alternate sources of energy, which as well as being beneficial to governments are also preferable to environmental campaigners. It has been suggested by some that politicians are providing scientists with funding to prove the $CO_2$ – temperature link and that the IPCC's findings are politically lead. Climate science is a big industry now with a lot of money invested and jobs depending on it.

By the end of the mid century, the Earth is expected to have 3 billion more people to feed and, because $CO_2$ is the main source of plant food, the easiest and most environmentally harmless way to increase food production would be to double levels of $CO_2$. This would have little effect on temperature (the $CO_2$- temperature forcing equation is logarithmic) but a significant effect on crop productivity, up to 40% in some varieties of soy bean. As the global temperatures fall, the oceans will absorb more $CO_2$ and $CO_2$ levels will begin to fall, this coupled with our efforts to try and cut $CO_2$ emissions will make world food production far more difficult. Because of this we must be very certain that $CO_2$ is causing a major change in the climate before we take action.

If cosmic rays are the main driver of climate change, it is good news for the world's inhabitants as it infers that the effect of carbon dioxide is quite small and, although there is nothing we can do about it, any global warming in the 21$^{st}$ century is likely to be much less than the typical predictions of 3 or 4°C.

It seems that the climate of our planet is such a complicated system that not enough is known to draw definite conclusions about the reasons for climate change. It does, however, seem difficult to believe that our species, that has dominated the planet for a relatively short period of time, could have such a huge impact on our planet's climate, whilst the Sun, the most massive body in the solar system whose influence dominates our planet, could have such little impact. Clearly all the causes discussed in here have an impact but their exact contributions have yet to be determined. Possibly one useful step forward could be made by a reassessment of all the data gathered concerning the Sun's activity since the days of William Herschel. This, together with the data gathered in other countries over hundreds of years, needs to be re-examined before possibly erroneous claims are made which could have disastrous consequences for mankind.

## Appendix A: IPCC language

[1] Numbers in square brackets indicate an IPCC calculated 90% uncertainty interval around a best estimate, i.e. there is an estimated 5% likelihood that the value could be above the range given in square brackets and 5% likelihood that the value could be below that range. Uncertainty intervals are not necessarily symmetric around the corresponding best estimate.

[2] The IPCC language of uncertainty

> **Box 1: The IPCC language of certainty**
> The IPCC uses the following italicised terms:
>
> *Virtually certain* >99% probability of occurrence; *Extremely likely* >95%; *Very likely* >90%; *Likely* >66%; *More likely than not* >50%; *Unlikely* <33%; *Very unlikely* <10%; *Extremely unlikely* <5%; *Exceptionally unlikely* <1%.